\documentclass[fleqn,usenatbib]{mnras}

\usepackage{amssymb,amsmath,framed}

\usepackage{bm}
\expandafter\ifx\csname package@font\endcsname\relax\else
 \expandafter\expandafter
 \expandafter\usepackage
 \expandafter\expandafter
 \expandafter{\csname package@font\endcsname}
\fi
\usepackage[T1]{fontenc}

\DeclareRobustCommand{\VAN}[3]{#2}
\let\VANthebibliography\thebibliography
\def\thebibliography{\DeclareRobustCommand{\VAN}[3]{##3}\VANthebibliography}

\usepackage{graphicx}
\usepackage{bm}
\usepackage{color}
\usepackage{newtxtext,newtxmath}

\title[Birth-death-migration model]{A Birth-Death-Migration Model for Life in Astrophysical Environments}

\author[Lingam et al.]{Manasvi Lingam$^{1}$\thanks{E-mail: mlingam@fit.edu}, Claudio Grimaldi$^{2,3}$ and Amedeo Balbi$^{4}$
\\
$^{1}$Department of Aerospace, Physics and Space Sciences, Florida Institute of Technology, Melbourne FL 32901, USA\\
$^{2}$Laboratory of Statistical Biophysics, Ecole Polytechnique F\'ed\'erale de Lausanne - EPFL, 1015 Lausanne, Switzerland\\
$^{3}$Centro Studi e Ricerche Enrico Fermi, 00184 Roma, Italy\\
$^{4}$Dipartimento di Fisica, Universit\`a di Roma ``Tor Vergata", 00133 Roma, Italy \\
}
\date{Accepted XXX. Received YYY; in original form ZZZ}

\pubyear{2021}

\begin{document}
\label{firstpage}
\pagerange{\pageref{firstpage}--\pageref{lastpage}}
\maketitle

\begin{abstract}
To assess the number of life-bearing worlds in astrophysical environments, it is necessary to take the intertwined processes of abiogenesis (birth), extinction (death), and transfer of life (migration) into account. We construct a mathematical model that incorporates this trio of mechanisms and accordingly derive the probability distribution function and other statistical properties (e.g., mean) for the number of worlds with biospheres. We show that a given astrophysical setting may become eventually saturated with life if the rate of successful transfers of organisms is higher than the extinction rate of biospheres. Based on the available data, we suggest that this criterion might be fulfilled for star-forming clusters (and perhaps the Galactic bulge under optimal circumstances), thereby indicating that such regions could constitute promising abodes for hosting and detecting life. 
\end{abstract}




\section{Introduction}\label{SecIntro}
At their very heart, attempts to model the distribution of life-bearing worlds in astrophysical environments must take two processes into account. On the one hand, new life-bearing worlds burst onto the scene, on which abiogenesis has facilitated the emergence of life \emph{de novo}. On the other, a panoply of phenomena, ranging from the astrophysical to the geological, can instigate the extinction of biospheres, thereby diminishing the number of life-bearing worlds. Hence, if we take merely these two mechanisms into consideration, we enter the realm of birth-death models, which are unsurprisingly widely employed in the natural sciences \citep{Yu24,Bai64,MSB78,Nee06,VK07}.

However, there is one more crucial process that is oft-neglected: life could migrate from one world to another, loosely akin to the migration of species between islands in biogeography \citep{MW67}. On the scale of planetary systems, this mechanism of panspermia \citep{Melosh1988, Wes10,Wick2010,ML21}, as it is known, has a rich history dating back to the ancient Greeks and further beyond in time. This factor is important because at least one specific instantiation of panspermia has been argued to have some chances of operating over interstellar distances -- to wit, ``lithopanspermia'', whereby lifeforms embedded in rocks are expelled from one planetary system and captured by others \citep{Zubrin2001,Wallis2004,Napier2004,ML16,GLB18,Siraj2020,GHSH}. Interstellar (litho)panspermia might be especially effective in crowded environments, such as in star-forming clusters \citep{AS05,Valtonen2009,BMM12} or in the Galactic bulge \citep{CFL18,BHK20,GHSH}, as explicated hereafter.

Thus, if one wishes to properly model the distribution of life-bearing worlds, it is necessary to encompass the process of migration as well, which necessitates developing a suitable birth-death-migration model to analyse the resultant dynamics and how that engenders an increase or decrease in the number of biospheres. This constitutes the chief objective of the paper and represents a departure from prior studies. Thus, in our approach, the probability that life appears on a suitable (i.e., habitable) set of planets is not simply governed by endogenous abiogenesis, but also by the likelihood that living organisms which had originated elsewhere can migrate safely and gain a firm foothold in the target location. 

The preceding statement could be reframed along the following lines. It is common to view abiogenesis and the migration of life (viz., panspermia) as being in tension with each other. When it comes to a \emph{specific} planet, life on that world would have been initiated by either abiogenesis or panspermia, which explains why they are often perceived as competing hypotheses. On the other hand, if we broaden our horizon so to speak, the two processes are arguably complementary and two sides of the same proverbial coin (see \citealt{GH07}), in the sense that both of them may act to enhance the number of life-bearing worlds in a particular astrophysical environment. It is, therefore, instructive to analyse the efficacy of both these mechanisms (in a given setting) in tandem, which is the \emph{summum bonum} of this paper. 

Our framework has two other interesting facets that merit a mention. First, it relies solely on the ratios of timescales of the competing mechanisms: while these timescales are admittedly as yet unknown, their relative importance can nevertheless be estimated for specific situations in a semiquantitative fashion, therefore furnishing some insight into the prevalence of inhabited worlds in different astrophysical environments. Second, as our model emphasises the temporal aspect of the processes involved, it could shed some light on how the number of life-bearing worlds evolves over the course of the history of the astrophysical environment in question.

At the outset, we wish to caution that our model examines the dynamical facets of lithopanspermia. To put it differently, it does not delve deeply into the biological realm, where there are substantive uncertainties and unknowns about the long-term survival of organisms embedded in rocks as well as their capability to withstand the ejection and entry phases \citep{MB04}. Hence, the criteria we derive for the effectuation of panspermia must be perceived, at best, as necessary, but not sufficient, conditions.

The outline of the paper is as follows. In Section \ref{SecMod}, we present the assumptions underpinning our quantitative model and derive the ensuing results, while explicating the attendant caveats. In Section \ref{SecDisc}, we apply the formalism to specific astrophysical environments and outline the ramifications of our analysis.

\section{Mathematical model}\label{SecMod}
Clearly, the core issue of attempting to estimate the number of life-bearing worlds in a chosen astrophysical environment (denoted by $n$) is hugely complex and requires careful handling.

One possible approach is to adopt a formalism akin to the Drake equation \citep{Drake65}, but there are a minimum of two major drawbacks that arise: (1) any formalism similar to the Drake equation will run into a plethora of unknown parameters, and (2) most of these parameters are not precisely delineated and are specified as point estimates \citep{Ci04}. Hence, to bypass these obstacles, several studies have sought to recast the Drake equation in statistical terms \citep{Mac10,GBB12}. In particular, the recent studies by \citet{CJFY} and \citet{Kip21} modelled the number of technological species by assigning typical timescales (or equivalently rates) for their origin, evolution, and death. In a similar vein, the framework of cellular automata was utilised by \citet{DVC19} to analyse habitability on Galactic scales by accounting for the twin mechanisms of expansion and extinction. 

If we consider worlds with non-technological life, it is apparent that a similar procedure can be harnessed to study the value of $n$. Before doing so, we highlight what is perhaps the most crucial caveat of this analysis. The variables employed in our model (viz., the rates) should be implicitly understood to represent temporal averages. In actuality, $n$ may fluctuate over time as a result of variations in the frequency of phenomena affecting habitability on interstellar scales, such as supernovae \citep{MT11,BW17}, gamma-ray bursts \citep{Dart11,PJ14}, active galactic nuclei \citep{BT17,FL18}, and tidal disruption events \citep{PBL20}. Furthermore, instead of focusing on $n$, if we were to contemplate the number density of life-bearing worlds, that quantity is also expected to be heterogeneous, and may attain maximal values in the so-called Galactic Habitable Zone \citep{LFG04,Pra08} or even near the bulge of spiral galaxies \citep{BHK20}.

Bearing this central proviso in mind, let us consider the dynamical mechanisms whereby $n$ changes with time. 
\begin{enumerate}
    \item Independent abiogenesis events will increase the number of biospheres (i.e., life-bearing worlds). By positing that the constant rate of such events is $k_A$, we may express the result as $n \xrightarrow{{k_A}} n+1$. 
    \item Biospheres can become extinct due to a number of factors ranging from the geological to the planetary (e.g., asteroid or comet impacts) and galactic (e.g., supernovae). If biospheres become extinct at the rate $k_E$ per life-bearing world (which would translate to a total extinction rate of $k_E \cdot n$, because there are $n$ such worlds at a given time), the process can be expressed as $n \xrightarrow{{k_E \cdot n}} n-1$.
    \item In a successful instantiation of panspermia, one preexisting world gives ``birth'' to two worlds, analogous to how fission operates in cells and organelles in biology \citep{ON03,Marg05}. Alternatively, we may envisage panspermia as being akin to the migration of species from a parent patch to an unoccupied patch in ecology.\footnote{Needless to say, these analogies with ecology should not be viewed as constituting exact correspondences.} Thus, if we assign a rate of $k_P$ per life-bearing world to this mechanism, the governing equation is tantamount to $n \xrightarrow{{k_P \cdot n}} n+1$; note that $k_P$ is representative of the rate of successful panspermia events.
\end{enumerate}
At this stage, a few comments are warranted. First, if we consider just (i) and (ii) above, the result is the classical birth-death process, which has been thoroughly investigated in several fields of physics, chemistry, and biology \citep{Yu24,MSB78,Nee06,VK07}. Second, (ii) and (iii) are widely prevalent in ecology, where they embody the phenomena of extinction and migration, respectively. In fact, these two processes constitute the foundation of classic ecological models such as the theory of island biogeography \citep{MW67} and metapopulations \citep{Lev69}. One might even envision panspermia, \emph{mutatis mutandis}, as island ecology on astronomical scales \citep{CBW07,Co08,LL17}. Third, the trio (i), (ii), and (iii) are encountered often in the scientific literature \citep{Bai64} -- to take just one example, they constitute a \emph{subset} of a generic biophysical model for ascertaining the number of organelles \citep{MOS14}.\footnote{The reason we describe them as a subset is because the biophysical model is also endowed with fusion, which is missing herein since two worlds with biospheres cannot `merge' to become just one life-bearing world.}

Moving on, we reiterate that the rates $k_A$, $k_E$ and $k_P$ are modelled as though they are constant over the spatial and temporal scales of the environment in question, i.e., we effectively consider the averaged values of these parameters. In actuality, these parameters will vary from one epoch to another -- for example, in the early Universe, with its higher rates of star formation and attendant high-energy phenomena such as gamma-ray bursts \citep[e.g.,][]{PJ14}, the rate of extinction ($k_E$) is anticipated to be higher. As we show hereafter in (\ref{MeanV}), the mean number of worlds hosting life (denoted by $\langle{n}\rangle$) is a function of $k_A$, $k_E$ and $k_P$. Hence, while $\langle{n}\rangle$ may attain some `characteristic' value when the suitable averages of $k_A$, $k_E$ and $k_P$ are substituted into the expression, there are likely to be fluctuations in $\langle{n}\rangle$ engendered by variations in $k_A$, $k_E$ and $k_P$. As our work is meant to develop a basic mathematical framework, we do not model the distribution of $n$ for spatially and temporally heterogeneous rates, which calls for detailed numerical modelling and is left as a vital subject for future work.

Lastly, it is worth appreciating that the three rates may exhibit some degree of correlation. To illustrate this scenario, consider the situation whereby close stellar encounters destabilise asteroid belts and boost the rates of impacts on to a planet bearing life. In that event, it is apparent that $k_E$ could be elevated (depending on the severity of the impacts) but this process can lead to the ejection of more debris from the planet, thereby potentially boosting $k_P$ as well. Hence, aside from high-energy radiation mentioned earlier, the role of stellar encounters and cognate perturbations (e.g., bolide impacts) must also be taken into account. As described in Section \ref{SecDisc}, close stellar encounters are rendered feasible, even likely, in the dense environment of the Galactic bulge.

The master equation of our model for the probability distribution function $P(n,t)$ for the specified astrophysical environment hosting $n$ life-bearing worlds at a given time $t$ is furnished below.
\begin{eqnarray}\label{MasEq}
  \frac{d P(n,t)}{d t} &=& \left[k_A + k_P (n - 1)\right] P(n-1,t) \nonumber \\
  && \,\, + k_E (n+1) P(n+1,t) \nonumber \\
  && \, - \left(k_A + k_P n + k_E n\right) P(n,t).
\end{eqnarray}
This master equation, which is a subset of \citet{MOS14} along expected lines, can be analysed to extract salient properties of the steady-state distribution function $P(n)$ such as the mean and variance \citep{Cra16}. Moreover, it is directly solvable to obtain the steady-state distribution function using the method of detailed balance \citep{VK07}. In what follows, we will make an important implicit assumption, namely, that the total number of potentially habitable worlds in the chosen astrophysical environment ($N_\mathrm{hab}$) is orders of magnitude larger than unity, which permits us to work with the limit of $N_\mathrm{hab} \rightarrow \infty$ for mathematical convenience, without sacrificing much accuracy.

For pedagogical reasons, we will briefly sketch how detailed balance may be employed to find $P(n)$ for the classical birth-death process comprising (i) and (ii) and thence generalised to encompass (iii); further details are provided in \citet{CDM19}. In the steady state, for the classical birth-death process, from (\ref{MasEq}) we have
\begin{equation}
  k_A  \, P(n-1) = k_E\, n\, P(n),
\end{equation}
which is a recursion relation for $P(n)$. After the repeated application of this recursion relation to write the final answer in terms of $P(0)$, we end up with
\begin{equation}\label{Pnrecur}
    P(n) = \frac{1}{n!} \left(\frac{k_A}{k_E}\right)^n P(0).
\end{equation}
To complete our derivation of $P(n)$, it is necessary to ascertain the value of $P(0)$, which is determined by appealing to the normalisation condition,
\begin{equation}
    \sum_{n = 0}^\infty P(n) = 1.
\end{equation}
We have started the summation at $n = 0$ instead of $n = 1$ because we are dealing with generic environments, some of which may not host any life-bearing worlds whatsoever. Substituting (\ref{Pnrecur}) into the above equation yields the ensuing relations:
\begin{equation}
    P(0) = \exp\left(-\frac{k_A}{k_E}\right),
\end{equation}
\begin{equation}\label{PoissDis}
     P(n) = \frac{1}{n!} \left(\frac{k_A}{k_E}\right)^n \exp\left(-\frac{k_A}{k_E}\right).
\end{equation}
Thus, the probability distribution function $P(n)$ is none other than a Poisson distribution with the shape parameter $\lambda = k_A/k_E$, which represents a standard result \citep{VK07}. When we incorporate panspermia (viz., mechanism (iii)) and repeat the procedure, the probability distribution $P(n)$ is determined to be
\begin{equation}\label{PDFMod}
    P(n) = \frac{1}{n!}\left(\frac{k_P}{k_E}\right)^n \frac{\Gamma\left(\frac{k_A}{k_P} + n\right)}{\Gamma\left(\frac{k_A}{k_P}\right)} \left(1 -\frac{k_P}{k_E}\right)^{k_A/k_P},
\end{equation}
where $\Gamma(x)$ denotes the gamma function. To simplify our notation hereafter, we introduce $\lambda_1 \equiv k_P/k_E$, $\lambda_2 \equiv k_A/k_P$ and $\lambda \equiv \lambda_1 \lambda_2 \equiv k_A/k_E$. 

\begin{figure*}
$$
\begin{array}{cc}
  \includegraphics[width=7.3cm]{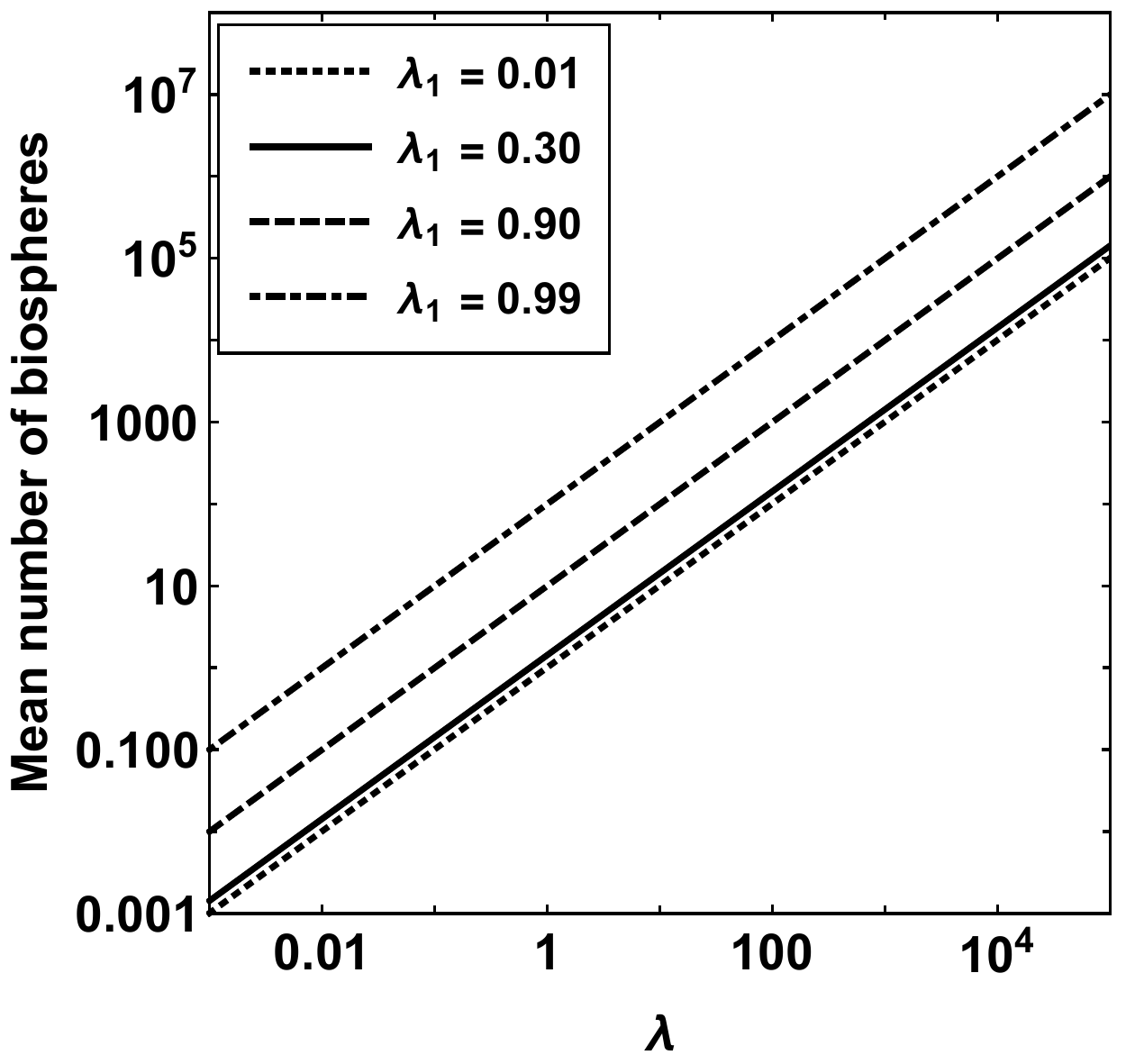} &  \includegraphics[width=7.4cm]{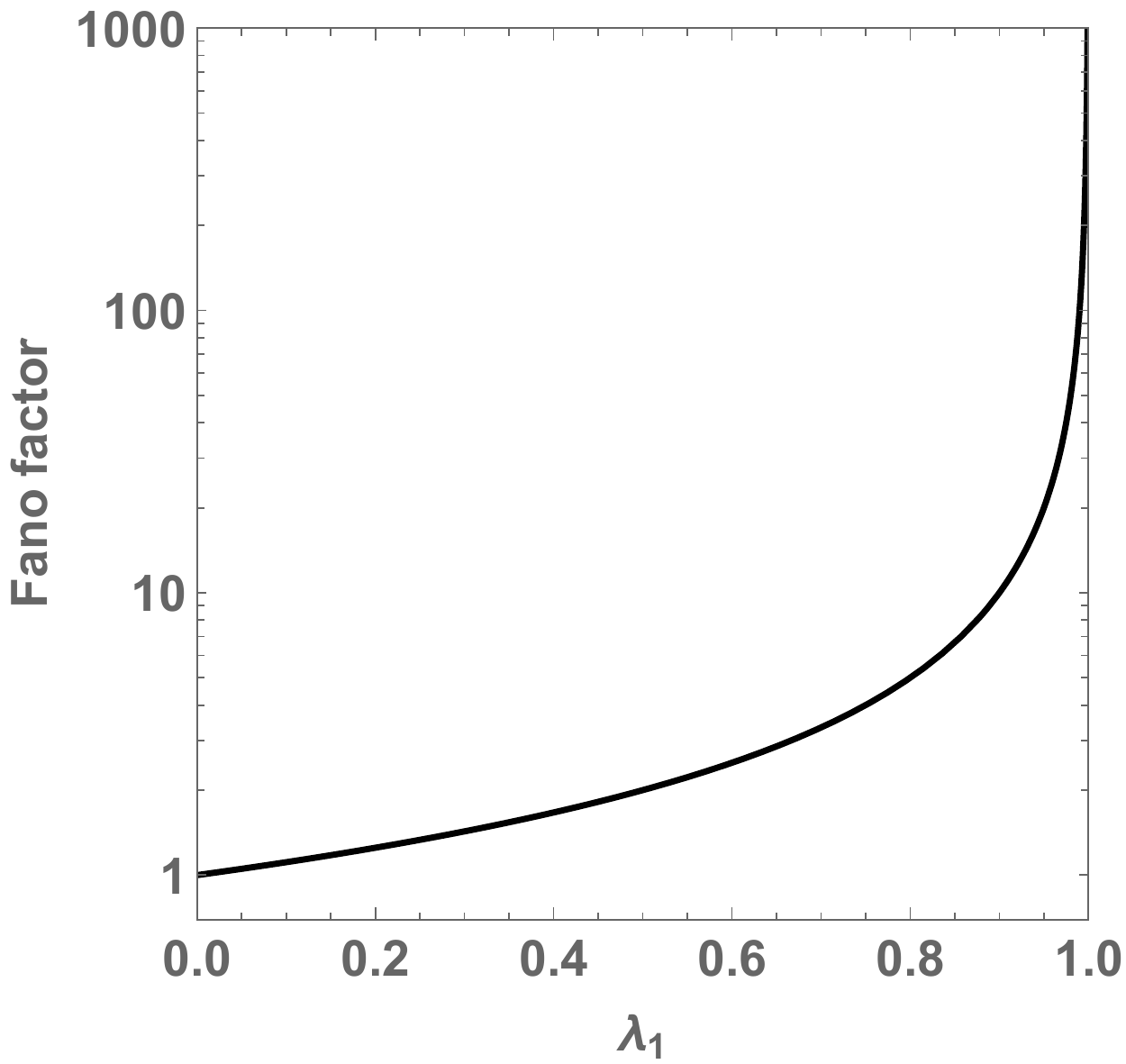}\\
\end{array}
$$
\caption{Left panel: average number of life-bearing worlds as a function of $\lambda$ (ratio of rates of abiogenesis and extinction) for different choices of $\lambda_1$ (ratio of rates of panspermia and extinction). Right panel: the Fano factor as a function of $\lambda_1$, the former of which quantifies the spread of the probability distribution.}
\label{FigMeanFano}
\end{figure*}

A number of interesting mathematical results follow from this distribution function. The mean number of life-bearing worlds $\langle{n}\rangle$ after calculation yields
\begin{equation}\label{MeanV}
    \langle{n}\rangle = \frac{\lambda}{1 - \lambda_1}.
\end{equation}
Now, suppose that we consider the limit wherein the rate of panspermia is much slower than that of extinction, which amounts to $\lambda_1 \ll 1$. In this limit, we find that $\langle{n}\rangle \rightarrow \lambda$, namely, the mean associated with the Poisson distribution of (\ref{PoissDis}). The interesting case, however, arises in the regime $\lambda_1 \approx 1$ because $\langle{n}\rangle \gg \lambda$ becomes feasible, as shown in Fig. \ref{FigMeanFano} (left panel) wherein we have plotted $\langle n\rangle$ as a function of $\lambda$ for different choices of $\lambda_1$. The third point worth noting is that (\ref{MeanV}) is meaningful (i.e., attains positive values) only when $k_P < k_E$ holds true -- to wit, extinction must be initiated more frequently than panspermia in order that a steady-state regime is reached. We note that a similar result was obtained by \citet{GLB21}, albeit through a simpler framework based on ordinary differential equations (ODEs). 

\begin{figure*}
\begin{center}
  \includegraphics[width=15cm]{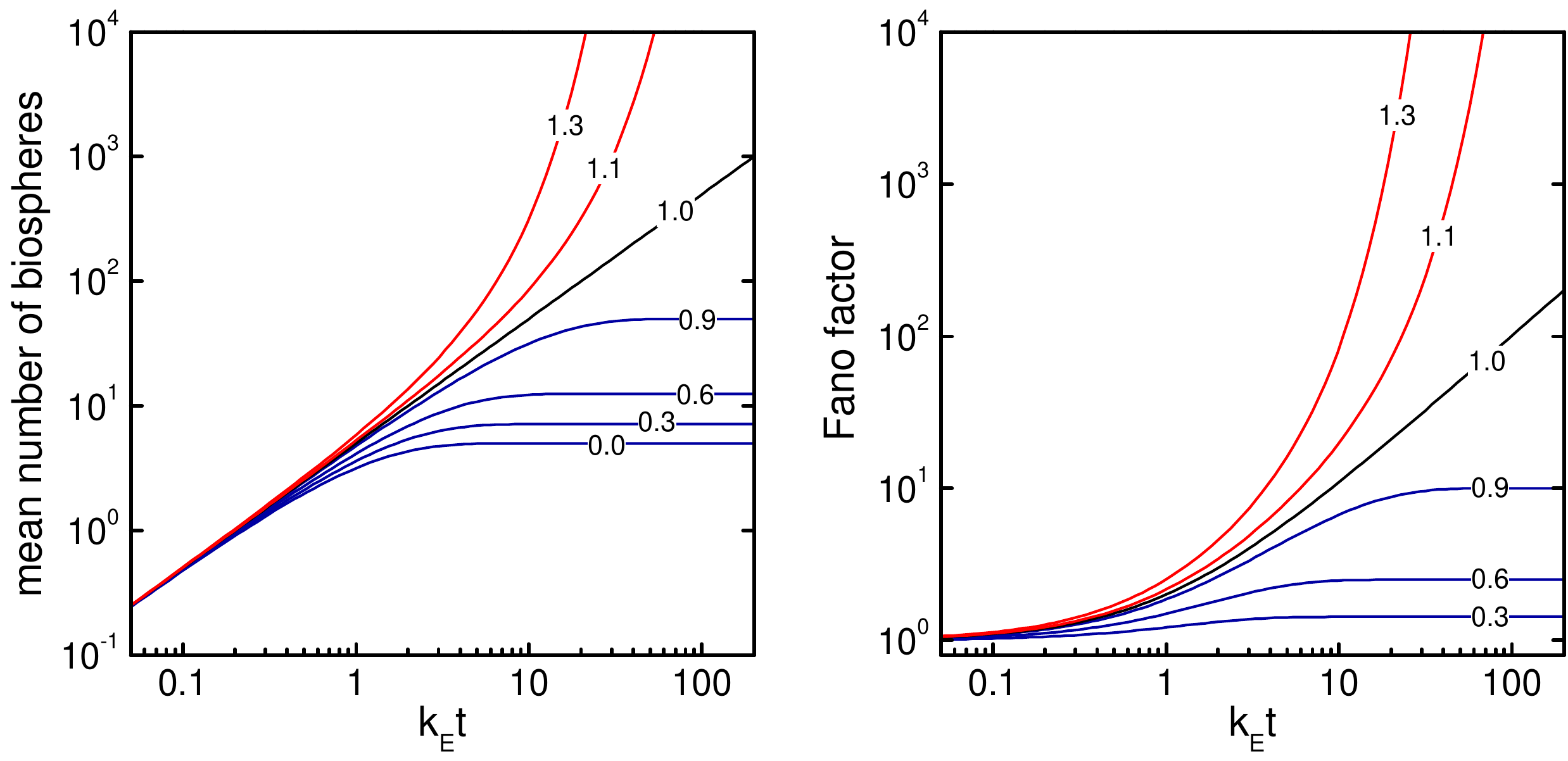} 
\caption{Left panel: average number of life-bearing worlds as a function of the dimensionless time variable $k_Et$ for $\lambda=5$ and different choices of $\lambda_1$. Right panel: the Fano factor as a function of $k_Et$ for different $\lambda_1$. The dark blue lines indicate the regime obtained for $\lambda_1<1$ in which  $\langle{n}\rangle$ and $\mathcal{F}_n$ asymptotically reach the steady-state values given in \eqref{MeanV} and \eqref{FanoF}, respectively. The red lines identify the solutions for $\lambda_1>1$, in which $\langle{n}\rangle$ and $\mathcal{F}_n$ increase exponentially with time. The initial condition has been chosen so that there are no biospheres at $t=0$.}
\label{FigTimeEvo}
\end{center}
\end{figure*}

Another quantity of interest to us is the Fano factor ($\mathcal{F}_n$), which is the variance divided by the mean and serves as a measure of the `width' of the probability distribution function. After carrying out the requisite calculations, we arrive at
\begin{equation}\label{FanoF}
    \mathcal{F}_n = \frac{1}{1 - \lambda_1}.
\end{equation}
By inspecting this expression, we notice once again that there is a singularity at $\lambda_1 = 1$, as shown in the right panel of Fig. \ref{FigMeanFano}, and $\mathcal{F}_n$ takes on non-physical values for $k_P > k_E$. Taken collectively, what these results imply is that the environment transitions from having a subset of potentially habitable worlds actually being populated with life at $\lambda_1 < 1$ to all such worlds hosting life in the limit of $\lambda_1 \geq 1$ given enough time. To make this point more manifest, we derive the differential equations for the temporal evolution of $\langle{n}\rangle$ and $\mathcal{F}_n$ from the master equation of \eqref{MasEq}, which is valid even for $k_P\geq k_E$. In terms of the dimensionless time variable $\tau=k_E t$, we obtain:
\begin{equation}\label{timen}
\frac{d \langle{n}\rangle}{d\tau}=\lambda-(1-\lambda_1)\langle{n}\rangle,
\end{equation}
\begin{equation}\label{timeF}
\frac{d\mathcal{F}_n}{d\tau}=1-(1-\lambda_1)\mathcal{F}_n,
\end{equation}   
whose solutions are plotted in Fig. \ref{FigTimeEvo} for different values of $\lambda_1$. From inspecting the figure and the preceding duo of equations, we see that the steady-state is attained when $(1-\lambda_1)\tau\gtrsim 1$ for $\lambda_1< 1$, which is tantamount to a timescale on the order of $t \gtrsim (k_E-k_P)^{-1}$, while at the critical point $\lambda_1=1$ both $\langle{n}\rangle$ and $\mathcal{F}_n$ increase linearly with time. Finally, if we consider $\lambda_1>1$, the mean number of biospheres and the corresponding Fano factor diverge as $t\rightarrow\infty$, indicating an exponentially increasing population of life-bearing worlds.

\begin{figure*}
$$
\begin{array}{ccc}
  \includegraphics[width=5.2cm]{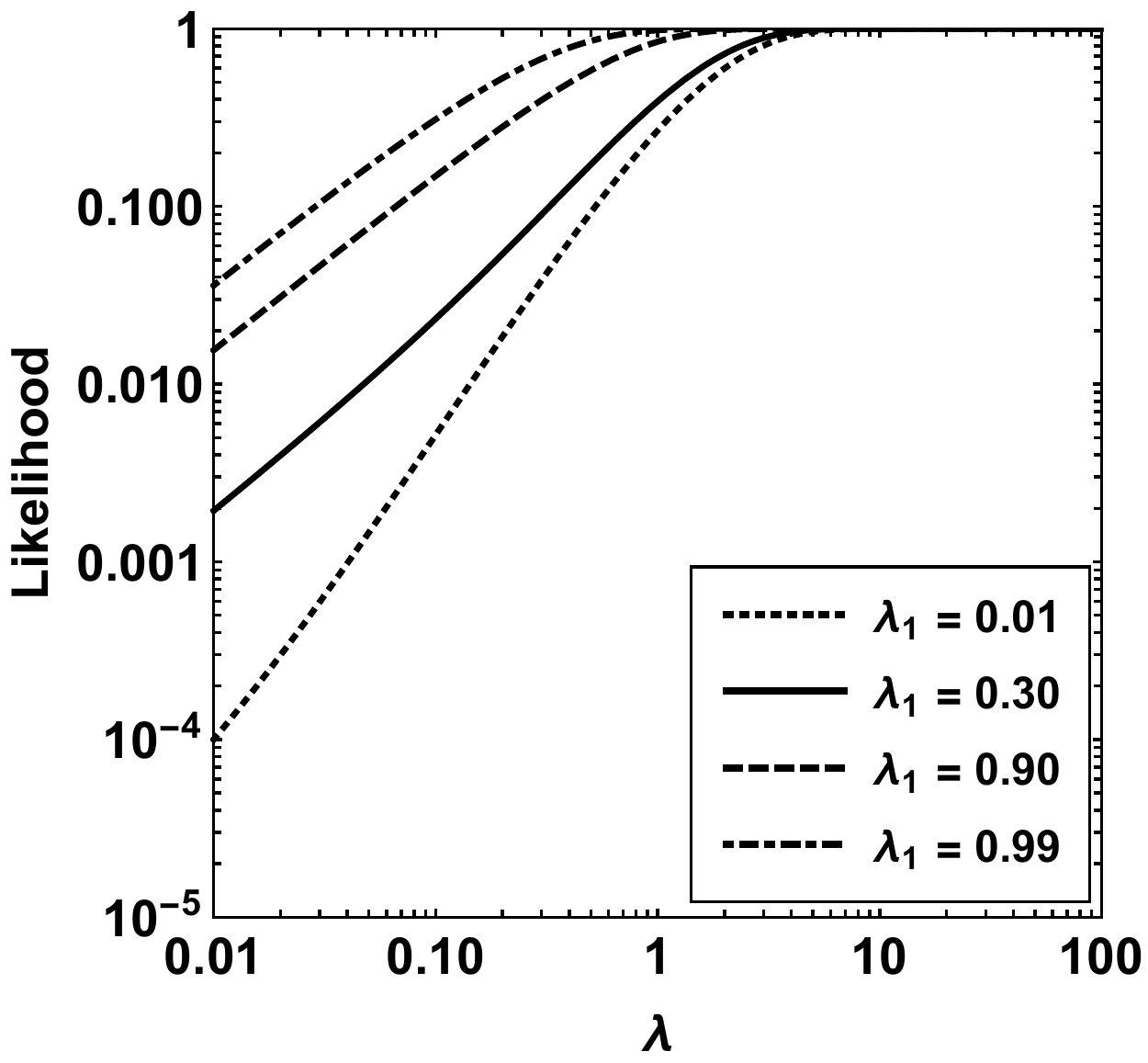} &  \includegraphics[width=5.0cm]{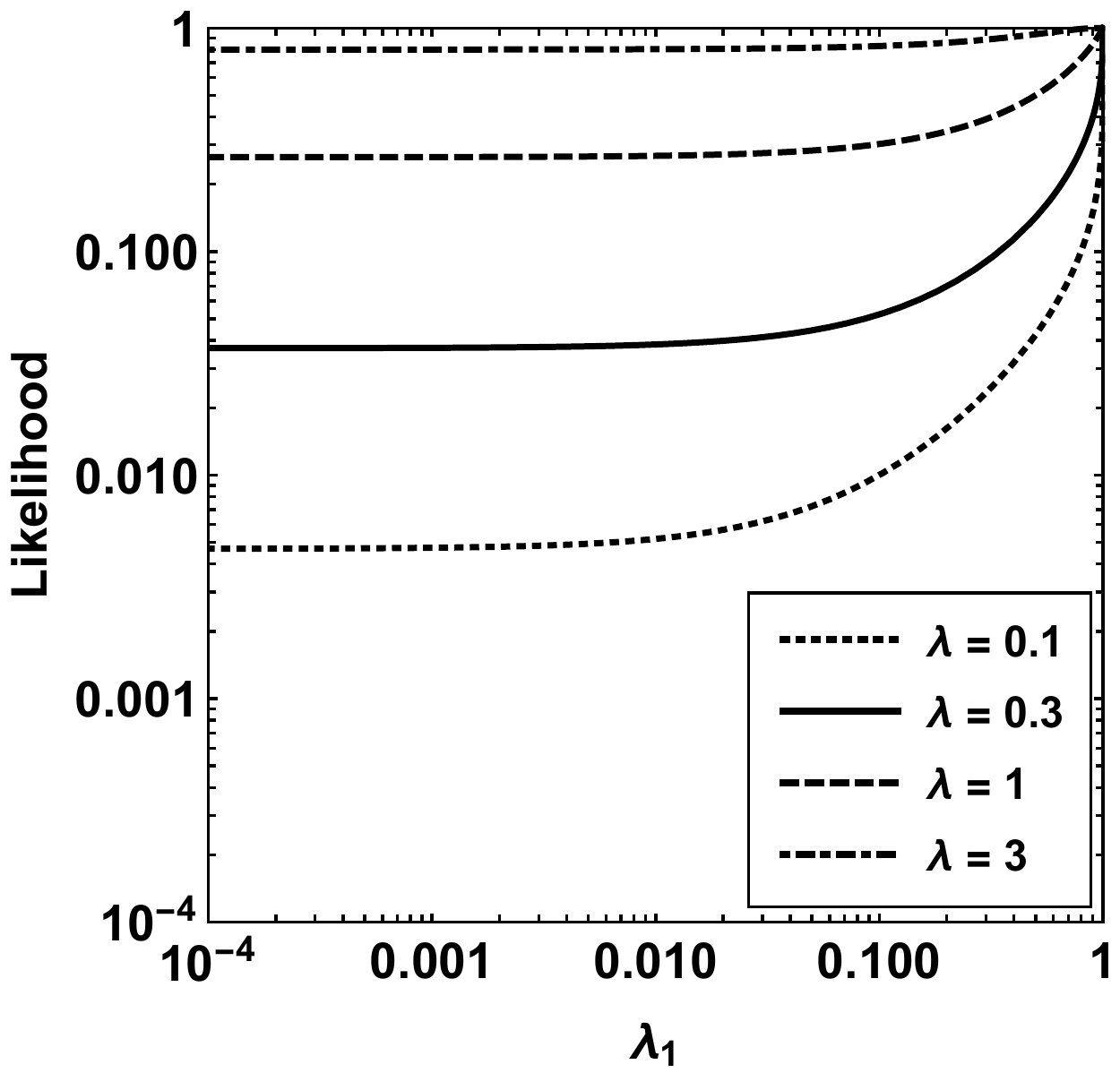} & \includegraphics[width=6.2cm]{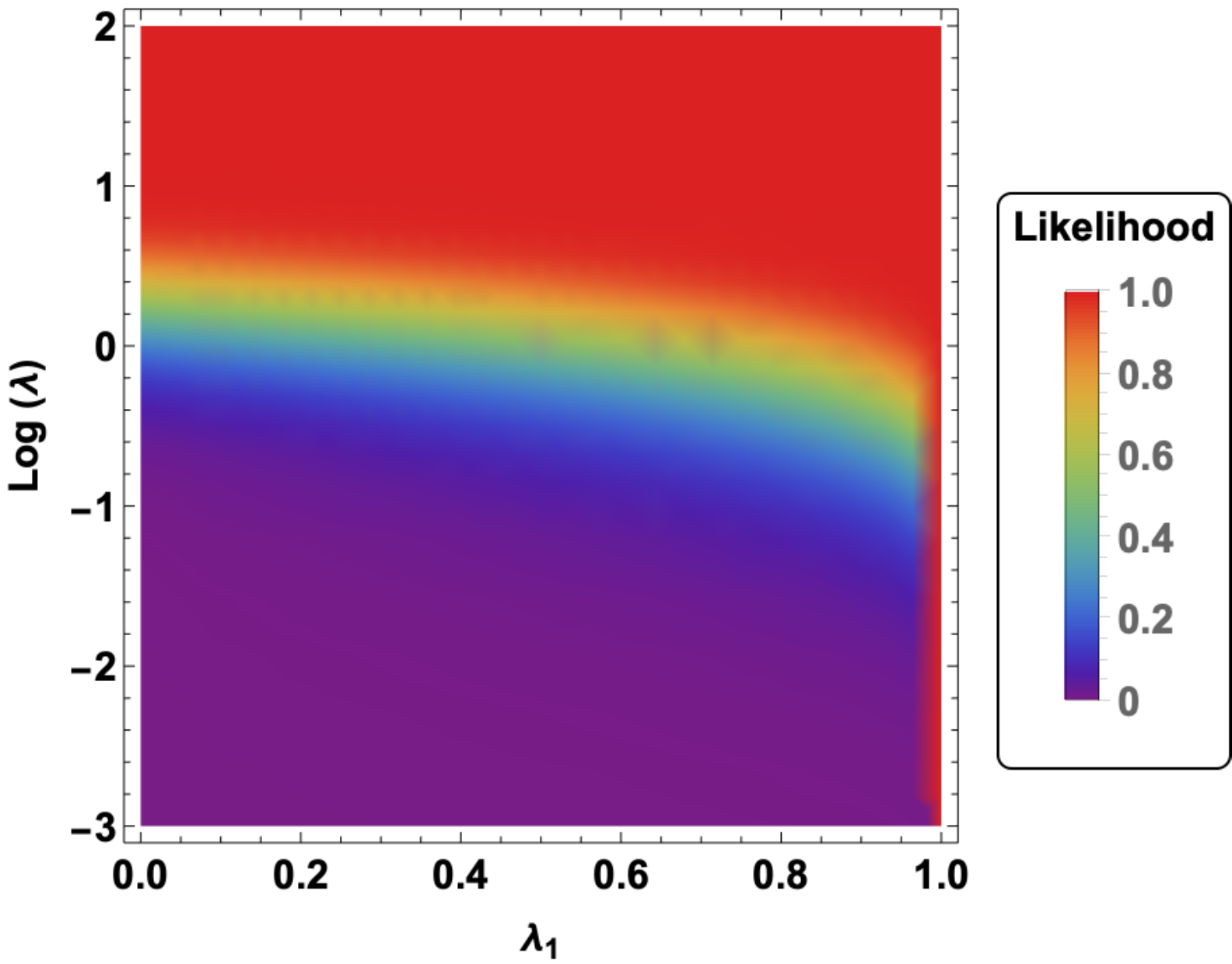}\\
\end{array}
$$
\caption{The likelihood that a particular astrophysical environments hosts more than one life-bearing world (such that no life-bearing world can be termed `alone') as a function of $\lambda$ (ratio of rates of abiogenesis and extinction) and $\lambda_1$ (ratio of rates of panspermia and extinction).}
\label{FigPNotAlone}
\end{figure*}

There is one more crucial quantity that should be calculated. In a given astrophysical environment, we are interested in the probability that a life-bearing world is not `alone', i.e., there are at least two worlds hosting life in the chosen setting. We denote this probability by $P_2$, and it is determined as follows:
\begin{equation}\label{P2def}
    P_2 = 1 - P(0) - P(1).
\end{equation}
After plugging in the appropriate values from (\ref{PDFMod}) and simplifying, we arrive at
\begin{equation}
    P_2 = 1 - \left(1 + \lambda\right) \left(1 - \lambda_1\right)^{\lambda/\lambda_1}
\end{equation}
We have plotted $P_2$ in Figure \ref{FigPNotAlone}, from which we can draw two interesting conclusions. First, in the case where $\lambda_1$ does not approach unity, we notice that there is a fairly sharp transition from $P_2 \ll 1$ to $P_2 \sim 1$ in the vicinity of $\lambda \sim 1$. Hence, when the rate of abiogenesis becomes comparable to that of extinction, the likelihood of finding another world hosting life increases substantially, which is along expected lines. Second, when $\lambda_1 \approx 1$ is valid, we notice that the transition to $P_2 \sim 1$ is actualised even if $\lambda$ is small. In other words, when the rate of panspermia is close to that of the extinction rate (or higher), panspermia can effectively populate the environment with life, thereby ensuring that the probability of a life-bearing world being alone is much reduced. 

\section{Discussion and Conclusions}\label{SecDisc}
Before embarking on our discussion, it is instructive to recapitulate our analysis, because we will draw on it hereafter. The key point is that, over time, an astrophysical environment transitions from being partly populated with life to being saturated with life when the critical threshold of $\lambda_1 = 1$ is crossed. Therefore, $k_P \geq k_E$ is the condition of interest to us, viz., the rate of panspermia must exceed that of the rate of extinction of biospheres. The immediate and obvious question is: can the above condition be fulfilled? At the outset, we emphasise that neither $k_P$ nor $k_E$ are accurately known in any environment. With this caveat in mind, it is nevertheless instructive to contemplate some potential values. 

For starters, if we examine $k_E$, it is apparent that Earth's biosphere has existed for $\sim 4$ Gyr, which suggests that $k_E \sim 10^{-10}-10^{-9}$ yr$^{-1}$ may be conceivable if one resorts to the Copernican Principle, also called the Principle of Mediocrity, and further supposes that the inverse of the biosphere lifetime is a heuristic measure of $k_E$. One crucial issue, however, is that the Copernican Principle cannot be applied in a straightforward fashion across large spatial and temporal scales \citep{CB20}. To offer just two examples, the early Universe had an elevated frequency of sterilizing gamma-ray bursts \citep{PJ14} and the Galactic bulge has a higher prevalence of supernovae \citep{LFG04}. Moreover, the interval of habitability of some planets might be higher than that of Earth or lower -- as per some trajectories of Martian habitability, life might have existed for only $\mathcal{O}(100)$ Myr before dying out due to inhospitable conditions \citep{CC14}. 

In addition, the duration spent by Earth-analogs in the habitable zones \citep{Dole,KWR93} of Sun-like stars is $6$-$7$ Gyr \citep{RCO13,WT15}. Hence, for this class of planets and stars, it might be tenable to work with a lower bound of $k_E \sim 10^{-10}$ yr$^{-1}$, although the actual value of $k_E$ could be higher by virtue of factors such as bolide impacts, perturbations of (bio)geochemical cycles, and stellar activity. If one considers planets around M-dwarfs, the habitable zone is obviously much more long-lived \citep{RCO13}, but M-dwarf exoplanets are possibly susceptible to serious issues such as stellar wind erosion, stellar flares and coronal mass ejections, paucity of UV radiation, and tidal locking, to name a few \citep{Dole,LL19}. Planets around K-dwarfs might, however, have the optimal combination of long-lived habitable zones and reduced deleterious effects \citep{KWR93,HA14,CG16,ML17,LL18,ML19}. Overall, we choose to err on the side of caution and specify a range of $k_E \sim 10^{-9}-10^{-8}$ yr$^{-1}$, to wit, an extinction rate $\sim 1$-$2$ orders of magnitude higher than what one may naively guess from the prior statements.

Now, let us select our first environment, namely, the Galactic field. We focus on lithopanspermia (transport of life via rocks) henceforth in lieu of radiopanspermia (transport of life through grains driven by radiation pressure), as the latter poses severe challenges to lifeforms embedded in the grains \citep{NL83,Wes10}. Numerical modelling by \citet{HJM03} and \citet{AS05} indicates that the number of ejecta transferred from one planetary system to another in $1$ Gyr ranges between $10^{-4}$ and $0.1$, respectively; the divergence arises because the recipient stellar system was endowed with different architectures. Of the objects transferred, only a fraction of them will comprise viable populations of lifeforms that can seed the recipient world, which we denote by $f_\mathrm{bio}$. By assembling this data, we determine that $k_P \sim f_\mathrm{bio} \times \left(10^{-13}-10^{-10}\right)$ yr$^{-1}$.

Although our estimates are undoubtedly tentative, for reasons outlined previously, they nevertheless permit us to infer a potentially important conclusion. In the ambient Galactic field, since $f_\mathrm{bio} \leq 1$ by construction, it is found from earlier that $k_P \lesssim 10^{-13}-10^{-10}$ yr$^{-1}$. On comparing this range with the interval for $k_E$ specified above, it is apparent that $k_P \ll k_E$ is expected to hold true. In other words, in this setting, the rate of (litho)panspermia appears to be negligible in comparison to the rate of extinction, which suggests why it is unlikely that lithopanspermia constitutes a major mechanism in the context of significantly boosting the number of life-bearing worlds in the Galactic field.

However, there is a vital feature that we must now take into consideration. Observations have demonstrated that the majority of stars, including our Sun, are born in star-forming clusters and groups, which are anticipated to persist over timescales of $\mathcal{O}(100)$ Myr \citep{FCA10,SP13}. The fact that lithopanspermia is efficacious in these star-forming clusters was demonstrated by \citet{AS05} and further refined via detailed numerical simulations of the weak escape and capture mechanisms by \citet{BMM12}. As per the modelling results of \citet{BMM12}, ejecta are transferred between nearby worlds in the cluster via lithopanspermia at the rate of $\sim 100$ objects every $\sim 100$ Myr \citep[pp. 832-833]{ML21}. 

A brief interlude is necessary at this juncture. Since star-forming clusters disperse after an interval of $\mathcal{O}(100)$ Myr, as noted above, it is necessary for the following processes to operate in timescales of $\lesssim 100$ Myr, which is supported by numerical simulations and observations: (i) formation of terrestrial planets \citep{MLO12,RM20}, (ii) ejection of objects from the donor planetary system \citep{HJM03}, and (iii) capture into the recipient planetary system and collision with the recipient world \citep{HJM03}. And lastly, it is necessary for at least one world inside the cluster to host life in the aforementioned timescale. Although life on Earth required several $100$ Myr \emph{at most} \citep[Chapter 2]{ML21}, some scenarios indicate that abiogenesis might occur in $\lesssim 100$ Myr \citep{LM94,BBB20}. And even if abiogenesis does not transpire in time, life on a single world within the cluster could be seeded from outside, which was shown to be dynamically plausible by \citet{AS05}. \emph{In toto}, star-forming clusters are tenable environments for panspermia, as further expounded in \citet{AS05} and \citet{BMM12}.

Thus, by invoking the factor of $f_\mathrm{bio}$ from before along with the information in the preceding paragraphs, $k_P \sim f_\mathrm{bio} \times 10^{-6}$ yr$^{-1}$ is feasible if one calculates the rate by taking the (temporal) average. Let us compare this value against the prior estimate for $k_E$. On doing so, we determine that $f_\mathrm{bio} \gtrsim 0.001$-$0.01$ might suffice to ensure that the rate of panspermia is comparable or exceeds the rate of extinction, in which case the star-forming clusters may perhaps end up being efficiently populated with life. Thus, as long as one world inside the cluster possesses life, our model suggests that it is theoretically conceivable for a substantial fraction of the potentially habitable worlds within the star-forming cluster (before its dispersal) to acquire life by means of lithopanspermia.

We reiterate that one of the major unknowns is $f_\mathrm{bio}$, the fraction of ejecta that can actually instantiate life via lithopanspermia on the recipient world. However, based on our preliminary calculations, the possibility that $f_\mathrm{bio}$ could be as low as $\sim 10^{-3}$ and still permit lithopanspermia in star-forming clusters is encouraging, albeit by no means definitive. As underscored previously, another prominent unknown is $k_E$, the rate of extinction of biospheres. If it turns out that $k_E$ is high, this trend would pose a twofold detriment. To begin with, the effectiveness of panspermia would be suppressed, as $\lambda_1$ would become much smaller than unity. More importantly, a high value of $k_E$ would lead to a significant reduction of $\lambda$, and thereby ensure that not even abiogenesis may permit life to take root in the chosen environment. As intimated by (\ref{MeanV}), the mean number of life-bearing worlds is governed by the intricate interplay of the rates of abiogenesis, extinction, and panspermia.

Another environment that invites careful consideration is the inner portion of the Milky Way (namely, the `bulge') comprising regions within $\sim 2$ kpc from the Galactic centre. Traditionally, the bulge has been deemed less hospitable to life compared to the disc, mainly because of the high rates of potentially hazardous astrophysical events (e.g., nearby supernovae explosions), and the ostensibly detrimental effects of ionising radiation instigated by the presence of an active supermassive black hole \citep{BT17,PBL20}; see also \citet{LGB19}. However, while the rate of supernovae in the bulge at any given location is certainly at least an order of magnitude higher relative to that in the solar neighbourhood \citep{BHK20}, this is not in itself a showstopper for the presence of life. In fact, even in the bulge, \emph{complete} planetary sterilisation events caused by nearby supernovae explosions---requiring this phenomena to occur at a distance of less than $0.04$ pc, as per \cite{SAL17}---are predicted to be sufficiently infrequent so as to not alter substantially the value of $k_E$ estimated in the disk \citep{BHK20}. 

On the other hand, the high stellar density in the bulge, which is up to an order of magnitude higher than in the disk even quite far away from the Galactic centre and $2$-$3$ orders of magnitude larger in the inner regions \citep[Tables 3 and 5]{RRDP}, makes it a favourable environment for lithopanspermia \citep{GHSH}. In this environment, the higher prevalence of supernovae is probably not the limiting factor, as explained in the prior paragraph. However, one prominent phenomenon that can boost $k_E$ is the rate of close encounters of stellar systems in the bulge, and this could lead to disruption or destabilization of habitable/inhabited planets. Numerical simulations by \citet[pg. 2105]{MKJ20} suggest that $80$ percent of stars in the Galactic bulge have experienced close encounters at $< 1000$ AU at a mean rate of $\geq 1$ per Gyr.\footnote{It is important to recognize that the nature of stellar encounters is sensitive to the velocity dispersion \citep{JPLS13}. Furthermore, to gauge the severity of encounters, it is necessary to assess not just the distance of closest approach but also the interaction time -- namely, fast stellar flybys may be less disruptive compared to their slower analogues.} Hence, on the basis of this model, it is reasonable to specify $k_E \sim 10^{-9}-10^{-8}$ yr$^{-1}$, which happens to be precisely the range delineated earlier.

By employing an analytical model that incorporated the rate and mean velocity of objects ejected from a life-bearing system along with the appropriate stellar number density and capture cross-section of stellar systems, \citet[Figure 3]{BHK20} estimated that a single location could seed planets in the Galactic bulge over a span of $\sim 1$ Gyr. Another method which yields this timescale entails repeating the calculations in \citet[Section 3.2]{AS05}, but with the velocity dispersion \citep{ZUN17} and stellar number density \citep{RRDP} chosen to match that of the Galactic bulge. Therefore, the time-averaged rate of lithopanspermia can be accordingly expressed as $k_P \sim f_\mathrm{bio} \times 10^{-9}$ yr$^{-1}$. Thus, the Galactic bulge is poised intriguingly at the critical threshold of $k_P \sim k_E$ provided that $f_\mathrm{bio} \sim 1$ is fulfilled and the lower limit for $k_E$ is chosen. Hence, under the most optimal circumstances, the Galactic bulge might be a fairly conducive environment for the instantiation of panspermia; under more realistic assumptions, however, panspermia may not function as the major contributor in amplifying the distribution of life-bearing worlds over time.

Incidentally, what have we discussed so far is based on contemplating extinction events that cause the total eradication of life from an inhabited world. Partial extinction events which, although not responsible for creating lifeless worlds, would cause severe damage or a `reset' of the biosphere are certainly anticipated to be much more frequent in any given astrophysical environment compared to their counterparts that wholly sterilise planets. Although the total number of life-bearing worlds would be left unchanged in our analysis, such catastrophic occurrences may impact the distribution of planets hosting complex, and perhaps intelligent, life. 

In closing, we remark that the status quo outlined hitherto is radically altered for directed panspermia -- to wit, the deliberate seeding of worlds with life by advanced technological species \citep{CO73,GL21} -- even in the Galactic field. However, at the outset, we emphasise a crucial difference. While the success of lithopanspermia (discussed heretofore) is largely dependent on what may be termed `chance' events (e.g., ejection and capture of debris), directed panspermia entails purposeful and targeted activities by technological species. These activities could plausibly vary from one species to another -- for example, conveying biological material in passive `packages' versus deploying active probes (of the normal or self-replicating kind) -- and the complex social, cultural, economic, and engineering factors that dictate them patently fall outside the scope of this paper.

With this note of caution in mind, the average distance of $\sim 1$ pc between neighbouring stars can be traversed in $\sim 10^5$ yr even with chemical rockets made by humans. In fact, light sails powered by laser arrays \citep{RLF84}, or even high-energy astrophysical phenomena \citep{LinMa}, may reach speeds of $\sim 0.1\,c$ and thus are capable of traversing the entire Galaxy in $\sim 10^6$ yr. Moreover, via the joint implementation of proper shielding and deployment of sophisticated propulsion technologies, it seems feasible to achieve $f_\mathrm{bio} \approx 1$ and $k_P \gg k_E$, which would ostensibly imply that the entire Galaxy could be populated with life by directed panspermia. 

Hence, even a single technological species may explore (or, in this scenario, seed) the entire Milky Way in a timescale of $\sim 10$-$100$ Myr \citep{RB07,Lin16,CNF19}. The overarching critical unknown in this instance is, of course, the frequency of technological species that exist over this timescale and are committed to participating in such an enterprise. This issue evidently shares close connections with the (in)famous Fermi paradox, which we shall not delve into because it has already been thoroughly analysed in the literature \citep{MMC18,ML21}.

\section*{Data Availability Statement}
No new data were generated or analysed in support of this research.

\section*{Acknowledgements}
ML wishes to acknowledge the resources provided by the Harvard Library system, which were valuable for carrying out this work. The authors would like to thank the reviewers, whose constructive feedback was valuable in refining and improving the paper.

\bibliographystyle{mnras}
\bibliography{PanModel}

\begin{thebibliography}{}
\makeatletter
\relax
\def\mn@urlcharsother{\let\do\@makeother \do\$\do\&\do\#\do\^\do\_\do\%\do\~}
\def\mn@doi{\begingroup\mn@urlcharsother \@ifnextchar [ {\mn@doi@}
  {\mn@doi@[]}}
\def\mn@doi@[#1]#2{\def\@tempa{#1}\ifx\@tempa\@empty \href
  {http://dx.doi.org/#2} {doi:#2}\else \href {http://dx.doi.org/#2} {#1}\fi
  \endgroup}
\def\mn@eprint#1#2{\mn@eprint@#1:#2::\@nil}
\def\mn@eprint@arXiv#1{\href {http://arxiv.org/abs/#1} {{\tt arXiv:#1}}}
\def\mn@eprint@dblp#1{\href {http://dblp.uni-trier.de/rec/bibtex/#1.xml}
  {dblp:#1}}
\def\mn@eprint@#1:#2:#3:#4\@nil{\def\@tempa {#1}\def\@tempb {#2}\def\@tempc
  {#3}\ifx \@tempc \@empty \let \@tempc \@tempb \let \@tempb \@tempa \fi \ifx
  \@tempb \@empty \def\@tempb {arXiv}\fi \@ifundefined
  {mn@eprint@\@tempb}{\@tempb:\@tempc}{\expandafter \expandafter \csname
  mn@eprint@\@tempb\endcsname \expandafter{\@tempc}}}

\bibitem[\protect\citeauthoryear{{Adams}}{{Adams}}{2010}]{FCA10}
{Adams} F.~C.,  2010, \mn@doi [Annu. Rev. Astron. Astrophys.]
  {10.1146/annurev-astro-081309-130830}, \href
  {https://ui.adsabs.harvard.edu/abs/2010ARA&A..48...47A} {48, 47}

\bibitem[\protect\citeauthoryear{{Adams} \& {Spergel}}{{Adams} \&
  {Spergel}}{2005}]{AS05}
{Adams} F.~C.,  {Spergel} D.~N.,  2005, \mn@doi [Astrobiology]
  {10.1089/ast.2005.5.497}, \href
  {https://ui.adsabs.harvard.edu/abs/2005AsBio...5..497A} {5, 497}

\bibitem[\protect\citeauthoryear{{Bailey}}{{Bailey}}{1964}]{Bai64}
{Bailey} N. T.~J.,  1964, {The Elements of Stochastic Processes with
  Applications to the Natural Sciences}.
New York: John Wiley \& Sons

\bibitem[\protect\citeauthoryear{{Balbi} \& {Tombesi}}{{Balbi} \&
  {Tombesi}}{2017}]{BT17}
{Balbi} A.,  {Tombesi} F.,  2017, \mn@doi [Sci. Rep.]
  {10.1038/s41598-017-16110-0}, \href
  {https://ui.adsabs.harvard.edu/abs/2017NatSR...716626B} {7, 16626}

\bibitem[\protect\citeauthoryear{{Balbi}, {Hami}  \&
  {Kova{\v{c}}evi{\'c}}}{{Balbi} et~al.}{2020}]{BHK20}
{Balbi} A.,  {Hami} M.,   {Kova{\v{c}}evi{\'c}} A.~B.,  2020, \mn@doi [Life]
  {10.3390/life10080132}, \href
  {https://ui.adsabs.harvard.edu/abs/2020arXiv200801419B} {10, 132}

\bibitem[\protect\citeauthoryear{{Bartlett}}{{Bartlett}}{1978}]{MSB78}
{Bartlett} M.~S.,  1978, {An Introduction to Stochastic Processes: With Special
  Reference to Methods and Applications}, 3rd edn.
Cambridge: Cambridge University Press

\bibitem[\protect\citeauthoryear{{Belbruno}, {Moro-Mart{\'\i}n}, {Malhotra}  \&
  {Savransky}}{{Belbruno} et~al.}{2012}]{BMM12}
{Belbruno} E.,  {Moro-Mart{\'\i}n} A.,  {Malhotra} R.,   {Savransky} D.,  2012,
  \mn@doi [Astrobiology] {10.1089/ast.2012.0825}, \href
  {https://ui.adsabs.harvard.edu/abs/2012AsBio..12..754B} {12, 754}

\bibitem[\protect\citeauthoryear{{Benner} et~al.,}{{Benner}
  et~al.}{2020}]{BBB20}
{Benner} S.~A.,  et~al., 2020, \mn@doi [ChemSystemsChem]
  {10.1002/syst.201900035}, 2, e1900035

\bibitem[\protect\citeauthoryear{{Bj{\o}rk}}{{Bj{\o}rk}}{2007}]{RB07}
{Bj{\o}rk} R.,  2007, \mn@doi [Int. J. Astrobiol.] {10.1017/S1473550407003709},
  \href {https://ui.adsabs.harvard.edu/abs/2007IJAsB...6...89B} {6, 89}

\bibitem[\protect\citeauthoryear{{Branch} \& {Wheeler}}{{Branch} \&
  {Wheeler}}{2017}]{BW17}
{Branch} D.,  {Wheeler} J.~C.,  2017, {Supernova Explosions}.
Berlin: Springer-Verlag, \mn@doi{10.1007/978-3-662-55054-0}

\bibitem[\protect\citeauthoryear{{Burchell}}{{Burchell}}{2004}]{MB04}
{Burchell} M.~J.,  2004, \mn@doi [Int. J. Astrobiol.]
  {10.1017/S1473550404002113}, \href
  {https://ui.adsabs.harvard.edu/abs/2004IJAsB...3...73B} {3, 73}

\bibitem[\protect\citeauthoryear{{Cai}, {Jiang}, {Fahy}  \& {Yung}}{{Cai}
  et~al.}{2021}]{CJFY}
{Cai} X.,  {Jiang} J.~H.,  {Fahy} K.~A.,   {Yung} Y.~L.,  2021, \mn@doi
  [Galaxies] {10.3390/galaxies9010005}, \href
  {https://ui.adsabs.harvard.edu/abs/2021Galax...9....5C} {9, 5}

\bibitem[\protect\citeauthoryear{{Carroll-Nellenback}, {Frank}, {Wright}  \&
  {Scharf}}{{Carroll-Nellenback} et~al.}{2019}]{CNF19}
{Carroll-Nellenback} J.,  {Frank} A.,  {Wright} J.,   {Scharf} C.,  2019,
  \mn@doi [Astron. J.] {10.3847/1538-3881/ab31a3}, \href
  {https://ui.adsabs.harvard.edu/abs/2019AJ....158..117C} {158, 117}

\bibitem[\protect\citeauthoryear{{Chen}, {Forbes}  \& {Loeb}}{{Chen}
  et~al.}{2018}]{CFL18}
{Chen} H.,  {Forbes} J.~C.,   {Loeb} A.,  2018, \mn@doi [Astrophys. J. Lett.]
  {10.3847/2041-8213/aaab46}, \href
  {https://ui.adsabs.harvard.edu/abs/2018ApJ...855L...1C} {855, L1}

\bibitem[\protect\citeauthoryear{{Choubey}, {Das}  \& {Majumdar}}{{Choubey}
  et~al.}{2019}]{CDM19}
{Choubey} S.,  {Das} D.,   {Majumdar} S.,  2019, \mn@doi [Phys. Rev. E]
  {10.1103/PhysRevE.100.022405}, \href
  {https://ui.adsabs.harvard.edu/abs/2019PhRvE.100b2405C} {100, 022405}

\bibitem[\protect\citeauthoryear{{{\'C}irkovi{\'c}}}{{{\'C}irkovi{\'c}}}{2004}]{Ci04}
{{\'C}irkovi{\'c}} M.~M.,  2004, \mn@doi [Astrobiology]
  {10.1089/153110704323175160}, \href
  {https://ui.adsabs.harvard.edu/abs/2004AsBio...4..225C} {4, 225}

\bibitem[\protect\citeauthoryear{{{\'C}irkovi{\'c}}}{{{\'C}irkovi{\'c}}}{2018}]{MMC18}
{{\'C}irkovi{\'c}} M.~M.,  2018, The Great Silence: Science and Philosophy of
  Fermi's Paradox.
Oxford: Oxford University Press

\bibitem[\protect\citeauthoryear{{{\'C}irkovi{\'c}} \&
  {Balbi}}{{{\'C}irkovi{\'c}} \& {Balbi}}{2020}]{CB20}
{{\'C}irkovi{\'c}} M.~M.,  {Balbi} A.,  2020, \mn@doi [Int. J. Astrobiol.]
  {10.1017/S1473550419000223}, \href
  {https://ui.adsabs.harvard.edu/abs/2020IJAsB..19..101C} {19, 101}

\bibitem[\protect\citeauthoryear{{Cockell}}{{Cockell}}{2008}]{Co08}
{Cockell} C.~S.,  2008, \mn@doi [Orig. Life Evol. Biosph.]
  {10.1007/s11084-007-9112-3}, \href
  {https://ui.adsabs.harvard.edu/abs/2008OLEB...38...87C} {38, 87}

\bibitem[\protect\citeauthoryear{{Cockell}}{{Cockell}}{2014}]{CC14}
{Cockell} C.~S.,  2014, \mn@doi [Astrobiology] {10.1089/ast.2013.1106}, \href
  {https://ui.adsabs.harvard.edu/abs/2014AsBio..14..182C} {14, 182}

\bibitem[\protect\citeauthoryear{{Cockell} et~al.,}{{Cockell}
  et~al.}{2007}]{CBW07}
{Cockell} C.~S.,  et~al., 2007, \mn@doi [Astrobiology] {10.1089/ast.2006.0038},
  \href {https://ui.adsabs.harvard.edu/abs/2007AsBio...7....1C} {7, 1}

\bibitem[\protect\citeauthoryear{{Craven}}{{Craven}}{2016}]{Cra16}
{Craven} C.~J.,  2016, \mn@doi [eLife] {10.7554/eLife.10167.001}, 5, e10167

\bibitem[\protect\citeauthoryear{{Crick} \& {Orgel}}{{Crick} \&
  {Orgel}}{1973}]{CO73}
{Crick} F.~H.~C.,  {Orgel} L.~E.,  1973, \mn@doi [Icarus]
  {10.1016/0019-1035(73)90110-3}, \href
  {https://ui.adsabs.harvard.edu/abs/1973Icar...19..341C} {19, 341}

\bibitem[\protect\citeauthoryear{{Cuntz} \& {Guinan}}{{Cuntz} \&
  {Guinan}}{2016}]{CG16}
{Cuntz} M.,  {Guinan} E.~F.,  2016, \mn@doi [Astrophys. J.]
  {10.3847/0004-637X/827/1/79}, \href
  {https://ui.adsabs.harvard.edu/abs/2016ApJ...827...79C} {827, 79}

\bibitem[\protect\citeauthoryear{{Dartnell}}{{Dartnell}}{2011}]{Dart11}
{Dartnell} L.~R.,  2011, \mn@doi [Astrobiology] {10.1089/ast.2010.0528}, \href
  {https://ui.adsabs.harvard.edu/abs/2011AsBio..11..551D} {11, 551}

\bibitem[\protect\citeauthoryear{{Dole}}{{Dole}}{1964}]{Dole}
{Dole} S.~H.,  1964, {Habitable Planets For Man}.
New York: Blaisdell Pub.~Co.

\bibitem[\protect\citeauthoryear{{{D}o{\v{s}}ovi{\'c}}, {Vukoti{\'c}}  \&
  {{\'C}irkovi{\'c}}}{{{D}o{\v{s}}ovi{\'c}} et~al.}{2019}]{DVC19}
{{D}o{\v{s}}ovi{\'c}} V.,  {Vukoti{\'c}} B.,   {{\'C}irkovi{\'c}} M.~M.,  2019,
  \mn@doi [Astron. Astrophys.] {10.1051/0004-6361/201834588}, \href
  {https://ui.adsabs.harvard.edu/abs/2019A&A...625A..98D} {625, A98}

\bibitem[\protect\citeauthoryear{{Drake}}{{Drake}}{1965}]{Drake65}
{Drake} F.~D.,  1965, {The Radio Search for Intelligent Extraterrestrial Life}.
Oxford: Pergamon Press, pp 323--345

\bibitem[\protect\citeauthoryear{{Forbes} \& {Loeb}}{{Forbes} \&
  {Loeb}}{2018}]{FL18}
{Forbes} J.~C.,  {Loeb} A.,  2018, \mn@doi [Mon. Not. R. Astron. Soc.]
  {10.1093/mnras/sty1433}, \href
  {https://ui.adsabs.harvard.edu/abs/2018MNRAS.479..171F} {479, 171}

\bibitem[\protect\citeauthoryear{{Forward}}{{Forward}}{1984}]{RLF84}
{Forward} R.~L.,  1984, \mn@doi [J. Spacecraft Rockets] {10.2514/3.8632}, \href
  {https://ui.adsabs.harvard.edu/abs/1984JSpRo..21..187F} {21, 187}

\bibitem[\protect\citeauthoryear{{Ginsburg} \& {Lingam}}{{Ginsburg} \&
  {Lingam}}{2021}]{GL21}
{Ginsburg} I.,  {Lingam} M.,  2021, \mn@doi [Res. Notes AAS]
  {10.3847/2515-5172/ac0f5a}, \href
  {https://ui.adsabs.harvard.edu/abs/2021RNAAS...5..154G} {5, 154}

\bibitem[\protect\citeauthoryear{{Ginsburg}, {Lingam}  \& {Loeb}}{{Ginsburg}
  et~al.}{2018}]{GLB18}
{Ginsburg} I.,  {Lingam} M.,   {Loeb} A.,  2018, \mn@doi [Astrophys. J. Lett.]
  {10.3847/2041-8213/aaef2d}, \href
  {https://ui.adsabs.harvard.edu/abs/2018ApJ...868L..12G} {868, L12}

\bibitem[\protect\citeauthoryear{{Glade}, {Ballet}  \& {Bastien}}{{Glade}
  et~al.}{2012}]{GBB12}
{Glade} N.,  {Ballet} P.,   {Bastien} O.,  2012, \mn@doi [Int. J. Astrobiol.]
  {10.1017/S1473550411000413}, \href
  {https://ui.adsabs.harvard.edu/abs/2012IJAsB..11..103G} {11, 103}

\bibitem[\protect\citeauthoryear{{Gobat}, {Hong}, {Snaith}  \& {Hong}}{{Gobat}
  et~al.}{2021}]{GHSH}
{Gobat} R.,  {Hong} S.~E.,  {Snaith} O.,   {Hong} S.,  2021, Astrophys. J.,
  \href {https://ui.adsabs.harvard.edu/abs/2021arXiv210908926G} {p.
  arXiv:2109.08926}

\bibitem[\protect\citeauthoryear{{Gordon} \& {Hoover}}{{Gordon} \&
  {Hoover}}{2007}]{GH07}
{Gordon} R.,  {Hoover} R.~B.,  2007, in {Hoover} R.~B.,  {Levin} G.~V.,
  {Rozanov} A.~Y.,   {Davies} P. C.~W.,  eds,  Society of Photo-Optical
  Instrumentation Engineers (SPIE) Conference Series Vol. 6694, Instruments,
  Methods, and Missions for Astrobiology X. p. 669404,
  \mn@doi{10.1117/12.737041}

\bibitem[\protect\citeauthoryear{{Grimaldi}, {Lingam}  \& {Balbi}}{{Grimaldi}
  et~al.}{2021}]{GLB21}
{Grimaldi} C.,  {Lingam} M.,   {Balbi} A.,  2021, \mn@doi [Astron. J.]
  {10.3847/1538-3881/abfe61}, \href
  {https://ui.adsabs.harvard.edu/abs/2021AJ....162...23G} {162, 23}

\bibitem[\protect\citeauthoryear{{Heller} \& {Armstrong}}{{Heller} \&
  {Armstrong}}{2014}]{HA14}
{Heller} R.,  {Armstrong} J.,  2014, \mn@doi [Astrobiology]
  {10.1089/ast.2013.1088}, \href
  {https://ui.adsabs.harvard.edu/abs/2014AsBio..14...50H} {14, 50}

\bibitem[\protect\citeauthoryear{{Jim{\'e}nez-Torres}, {Pichardo}, {Lake}  \&
  {Segura}}{{Jim{\'e}nez-Torres} et~al.}{2013}]{JPLS13}
{Jim{\'e}nez-Torres} J.~J.,  {Pichardo} B.,  {Lake} G.,   {Segura} A.,  2013,
  \mn@doi [Astrobiology] {10.1089/ast.2012.0842}, \href
  {https://ui.adsabs.harvard.edu/abs/2013AsBio..13..491J} {13, 491}

\bibitem[\protect\citeauthoryear{{Kasting}, {Whitmire}  \&
  {Reynolds}}{{Kasting} et~al.}{1993}]{KWR93}
{Kasting} J.~F.,  {Whitmire} D.~P.,   {Reynolds} R.~T.,  1993, \mn@doi [Icarus]
  {10.1006/icar.1993.1010}, \href
  {https://ui.adsabs.harvard.edu/abs/1993Icar..101..108K} {101, 108}

\bibitem[\protect\citeauthoryear{{Kipping}}{{Kipping}}{2021}]{Kip21}
{Kipping} D.,  2021, \mn@doi [Res. Notes AAS] {10.3847/2515-5172/abeb7b}, \href
  {https://ui.adsabs.harvard.edu/abs/2021RNAAS...5...44K} {5, 44}

\bibitem[\protect\citeauthoryear{{Lazcano} \& {Miller}}{{Lazcano} \&
  {Miller}}{1994}]{LM94}
{Lazcano} A.,  {Miller} S.~L.,  1994, \mn@doi [J. Mol. Evol.]
  {10.1007/BF00160399}, \href
  {https://ui.adsabs.harvard.edu/abs/1994JMolE..39..546L} {39, 546}

\bibitem[\protect\citeauthoryear{{Levins}}{{Levins}}{1969}]{Lev69}
{Levins} R.,  1969, \mn@doi [Bull. Entomol. Soc. America]
  {10.1093/besa/15.3.237}, 15, 237

\bibitem[\protect\citeauthoryear{{Lineweaver}, {Fenner}  \&
  {Gibson}}{{Lineweaver} et~al.}{2004}]{LFG04}
{Lineweaver} C.~H.,  {Fenner} Y.,   {Gibson} B.~K.,  2004, \mn@doi [Science]
  {10.1126/science.1092322}, \href
  {https://ui.adsabs.harvard.edu/abs/2004Sci...303...59L} {303, 59}

\bibitem[\protect\citeauthoryear{{Lingam}}{{Lingam}}{2016a}]{Lin16}
{Lingam} M.,  2016a, \mn@doi [Astrobiology] {10.1089/ast.2015.1411}, \href
  {https://ui.adsabs.harvard.edu/abs/2016AsBio..16..418L} {16, 418}

\bibitem[\protect\citeauthoryear{{Lingam}}{{Lingam}}{2016b}]{ML16}
{Lingam} M.,  2016b, \mn@doi [Mon. Not. R. Astron. Soc.]
  {10.1093/mnras/stv2533}, \href
  {https://ui.adsabs.harvard.edu/abs/2016MNRAS.455.2792L} {455, 2792}

\bibitem[\protect\citeauthoryear{{Lingam} \& {Loeb}}{{Lingam} \&
  {Loeb}}{2017a}]{LL17}
{Lingam} M.,  {Loeb} A.,  2017a, \mn@doi [Proc. Natl. Acad. Sci.]
  {10.1073/pnas.1703517114}, \href
  {https://ui.adsabs.harvard.edu/abs/2017PNAS..114.6689L} {114, 6689}

\bibitem[\protect\citeauthoryear{{Lingam} \& {Loeb}}{{Lingam} \&
  {Loeb}}{2017b}]{ML17}
{Lingam} M.,  {Loeb} A.,  2017b, \mn@doi [Astrophys. J. Lett.]
  {10.3847/2041-8213/aa8860}, \href
  {https://ui.adsabs.harvard.edu/abs/2017ApJ...846L..21L} {846, L21}

\bibitem[\protect\citeauthoryear{{Lingam} \& {Loeb}}{{Lingam} \&
  {Loeb}}{2018}]{LL18}
{Lingam} M.,  {Loeb} A.,  2018, \mn@doi [J. Cosmol. Astropart. Phys.]
  {10.1088/1475-7516/2018/05/020}, \href
  {https://ui.adsabs.harvard.edu/abs/2018JCAP...05..020L} {2018, 020}

\bibitem[\protect\citeauthoryear{{Lingam} \& {Loeb}}{{Lingam} \&
  {Loeb}}{2019a}]{ML19}
{Lingam} M.,  {Loeb} A.,  2019a, \mn@doi [Int. J. Astrobiol.]
  {10.1017/S1473550419000016}, \href
  {https://ui.adsabs.harvard.edu/abs/2019IJAsB..18..527L} {18, 527}

\bibitem[\protect\citeauthoryear{{Lingam} \& {Loeb}}{{Lingam} \&
  {Loeb}}{2019b}]{LL19}
{Lingam} M.,  {Loeb} A.,  2019b, \mn@doi [Rev. Mod. Phys.]
  {10.1103/RevModPhys.91.021002}, \href
  {https://ui.adsabs.harvard.edu/abs/2019RvMP...91b1002L} {91, 021002}

\bibitem[\protect\citeauthoryear{{Lingam} \& {Loeb}}{{Lingam} \&
  {Loeb}}{2020}]{LinMa}
{Lingam} M.,  {Loeb} A.,  2020, \mn@doi [Astrophys. J.]
  {10.3847/1538-4357/ab7dc7}, \href
  {https://ui.adsabs.harvard.edu/abs/2020ApJ...894...36L} {894, 36}

\bibitem[\protect\citeauthoryear{{Lingam} \& {Loeb}}{{Lingam} \&
  {Loeb}}{2021}]{ML21}
{Lingam} M.,  {Loeb} A.,  2021, {Life in the Cosmos: From Biosignatures to
  Technosignatures}.
Cambridge: Harvard University Press, \url
  {https://www.hup.harvard.edu/catalog.php?isbn=9780674987579}

\bibitem[\protect\citeauthoryear{{Lingam}, {Ginsburg}  \& {Bialy}}{{Lingam}
  et~al.}{2019}]{LGB19}
{Lingam} M.,  {Ginsburg} I.,   {Bialy} S.,  2019, \mn@doi [Astrophys. J.]
  {10.3847/1538-4357/ab1b2f}, \href
  {https://ui.adsabs.harvard.edu/abs/2019ApJ...877...62L} {877, 62}

\bibitem[\protect\citeauthoryear{{MacArthur} \& {Wilson}}{{MacArthur} \&
  {Wilson}}{1967}]{MW67}
{MacArthur} R.~H.,  {Wilson} E.~O.,  1967, {The Theory of Island Biogeography}.
Monographs in Population Biology, Princeton: Princeton University Press

\bibitem[\protect\citeauthoryear{{Maccone}}{{Maccone}}{2010}]{Mac10}
{Maccone} C.,  2010, \mn@doi [Acta Astronaut.]
  {10.1016/j.actaastro.2010.05.003}, \href
  {https://ui.adsabs.harvard.edu/abs/2010AcAau..67.1366M} {67, 1366}

\bibitem[\protect\citeauthoryear{{Margolin}}{{Margolin}}{2005}]{Marg05}
{Margolin} W.,  2005, \mn@doi [Nat. Rev. Mol. Cell Biol.] {10.1038/nrm1745}, 6,
  862

\bibitem[\protect\citeauthoryear{{McTier}, {Kipping}  \& {Johnston}}{{McTier}
  et~al.}{2020}]{MKJ20}
{McTier} M. A.~S.,  {Kipping} D.~M.,   {Johnston} K.,  2020, \mn@doi [Mon. Not.
  R. Astron. Soc.] {10.1093/mnras/staa1232}, \href
  {https://ui.adsabs.harvard.edu/abs/2020MNRAS.495.2105M} {495, 2105}

\bibitem[\protect\citeauthoryear{{Melosh}}{{Melosh}}{1988}]{Melosh1988}
{Melosh} H.~J.,  1988, \mn@doi [Nature] {10.1038/332687a0}, \href
  {https://ui.adsabs.harvard.edu/abs/1988Natur.332..687M} {332, 687}

\bibitem[\protect\citeauthoryear{{Melosh}}{{Melosh}}{2003}]{HJM03}
{Melosh} H.~J.,  2003, \mn@doi [Astrobiology] {10.1089/153110703321632525},
  \href {https://ui.adsabs.harvard.edu/abs/2003AsBio...3..207M} {3, 207}

\bibitem[\protect\citeauthoryear{{Melott} \& {Thomas}}{{Melott} \&
  {Thomas}}{2011}]{MT11}
{Melott} A.~L.,  {Thomas} B.~C.,  2011, \mn@doi [Astrobiology]
  {10.1089/ast.2010.0603}, \href
  {https://ui.adsabs.harvard.edu/abs/2011AsBio..11..343M} {11, 343}

\bibitem[\protect\citeauthoryear{{Morbidelli}, {Lunine}, {O'Brien}, {Raymond}
  \& {Walsh}}{{Morbidelli} et~al.}{2012}]{MLO12}
{Morbidelli} A.,  {Lunine} J.~I.,  {O'Brien} D.~P.,  {Raymond} S.~N.,   {Walsh}
  K.~J.,  2012, \mn@doi [Annu. Rev. Earth Planet. Sci.]
  {10.1146/annurev-earth-042711-105319}, \href
  {https://ui.adsabs.harvard.edu/abs/2012AREPS..40..251M} {40, 251}

\bibitem[\protect\citeauthoryear{{Mukherji} \& {O'Shea}}{{Mukherji} \&
  {O'Shea}}{2014}]{MOS14}
{Mukherji} S.,  {O'Shea} E.~K.,  2014, \mn@doi [eLife]
  {10.7554/eLife.02678.001}, 3, e02678

\bibitem[\protect\citeauthoryear{{Napier}}{{Napier}}{2004}]{Napier2004}
{Napier} W.~M.,  2004, \mn@doi [Mon. Not. R. Astron. Soc.]
  {10.1111/j.1365-2966.2004.07287.x}, \href
  {https://ui.adsabs.harvard.edu/abs/2004MNRAS.348...46N} {348, 46}

\bibitem[\protect\citeauthoryear{{Nee}}{{Nee}}{2006}]{Nee06}
{Nee} S.,  2006, \mn@doi [Annu. Rev. Ecol. Evol. Syst.]
  {10.1146/annurev.ecolsys.37.091305.110035}, 37, 1

\bibitem[\protect\citeauthoryear{{Nussinov} \& {Lysenko}}{{Nussinov} \&
  {Lysenko}}{1983}]{NL83}
{Nussinov} M.~D.,  {Lysenko} S.~V.,  1983, \mn@doi [Orig. Life]
  {10.1007/BF00928893}, \href
  {https://ui.adsabs.harvard.edu/abs/1983OrLi...13..153N} {13, 153}

\bibitem[\protect\citeauthoryear{{Osteryoung} \& {Nunnari}}{{Osteryoung} \&
  {Nunnari}}{2003}]{ON03}
{Osteryoung} K.~W.,  {Nunnari} J.,  2003, \mn@doi [Science]
  {10.1126/science.1082192}, \href
  {https://ui.adsabs.harvard.edu/abs/2003Sci...302.1698O} {302, 1698}

\bibitem[\protect\citeauthoryear{{Pacetti}, {Balbi}, {Lingam}, {Tombesi}  \&
  {Perlman}}{{Pacetti} et~al.}{2020}]{PBL20}
{Pacetti} E.,  {Balbi} A.,  {Lingam} M.,  {Tombesi} F.,   {Perlman} E.,  2020,
  \mn@doi [Mon. Not. R. Astron. Soc.] {10.1093/mnras/staa2535}, \href
  {https://ui.adsabs.harvard.edu/abs/2020MNRAS.498.3153P} {498, 3153}

\bibitem[\protect\citeauthoryear{{Pfalzner}}{{Pfalzner}}{2013}]{SP13}
{Pfalzner} S.,  2013, \mn@doi [Astron. Astrophys.]
  {10.1051/0004-6361/201218792}, \href
  {https://ui.adsabs.harvard.edu/abs/2013A&A...549A..82P} {549, A82}

\bibitem[\protect\citeauthoryear{{Piran} \& {Jimenez}}{{Piran} \&
  {Jimenez}}{2014}]{PJ14}
{Piran} T.,  {Jimenez} R.,  2014, \mn@doi [Phys. Rev. Lett.]
  {10.1103/PhysRevLett.113.231102}, \href
  {https://ui.adsabs.harvard.edu/abs/2014PhRvL.113w1102P} {113, 231102}

\bibitem[\protect\citeauthoryear{{Prantzos}}{{Prantzos}}{2008}]{Pra08}
{Prantzos} N.,  2008, \mn@doi [Space Sci. Rev.] {10.1007/s11214-007-9236-9},
  \href {https://ui.adsabs.harvard.edu/abs/2008SSRv..135..313P} {135, 313}

\bibitem[\protect\citeauthoryear{{Raymond} \& {Morbidelli}}{{Raymond} \&
  {Morbidelli}}{2020}]{RM20}
{Raymond} S.~N.,  {Morbidelli} A.,  2020, arXiv e-prints, \href
  {https://ui.adsabs.harvard.edu/abs/2020arXiv200205756R} {p. arXiv:2002.05756}

\bibitem[\protect\citeauthoryear{{Robin}, {Reyl{\'e}}, {Derri{\`e}re}  \&
  {Picaud}}{{Robin} et~al.}{2003}]{RRDP}
{Robin} A.~C.,  {Reyl{\'e}} C.,  {Derri{\`e}re} S.,   {Picaud} S.,  2003,
  \mn@doi [Astron. Astrophys.] {10.1051/0004-6361:20031117}, \href
  {https://ui.adsabs.harvard.edu/abs/2003A&A...409..523R} {409, 523}

\bibitem[\protect\citeauthoryear{{Rushby}, {Claire}, {Osborn}  \&
  {Watson}}{{Rushby} et~al.}{2013}]{RCO13}
{Rushby} A.~J.,  {Claire} M.~W.,  {Osborn} H.,   {Watson} A.~J.,  2013, \mn@doi
  [Astrobiology] {10.1089/ast.2012.0938}, \href
  {https://ui.adsabs.harvard.edu/abs/2013AsBio..13..833R} {13, 833}

\bibitem[\protect\citeauthoryear{{Siraj} \& {Loeb}}{{Siraj} \&
  {Loeb}}{2020}]{Siraj2020}
{Siraj} A.,  {Loeb} A.,  2020, \mn@doi [Life] {10.3390/life10040044}, 10, 44

\bibitem[\protect\citeauthoryear{{Sloan}, {Alves Batista}  \& {Loeb}}{{Sloan}
  et~al.}{2017}]{SAL17}
{Sloan} D.,  {Alves Batista} R.,   {Loeb} A.,  2017, \mn@doi [Sci. Rep.]
  {10.1038/s41598-017-05796-x}, \href
  {https://ui.adsabs.harvard.edu/abs/2017NatSR...7.5419S} {7, 5419}

\bibitem[\protect\citeauthoryear{{Valtonen} et~al.,}{{Valtonen}
  et~al.}{2009}]{Valtonen2009}
{Valtonen} M.,  et~al., 2009, \mn@doi [Astrophys. J.]
  {10.1088/0004-637X/690/1/210}, \href
  {https://ui.adsabs.harvard.edu/abs/2009ApJ...690..210V} {690, 210}

\bibitem[\protect\citeauthoryear{{Van Kampen}}{{Van Kampen}}{2007}]{VK07}
{Van Kampen} N.~G.,  2007, {Stochastic Processes in Physics and Chemistry}, 3rd
  edn.
Amsterdam: Elsevier

\bibitem[\protect\citeauthoryear{{Wallis} \& {Wickramasinghe}}{{Wallis} \&
  {Wickramasinghe}}{2004}]{Wallis2004}
{Wallis} M.~K.,  {Wickramasinghe} N.~C.,  2004, \mn@doi [Mon. Not. R. Astron.
  Soc.] {10.1111/j.1365-2966.2004.07355.x}, \href
  {https://ui.adsabs.harvard.edu/abs/2004MNRAS.348...52W} {348, 52}

\bibitem[\protect\citeauthoryear{{Wesson}}{{Wesson}}{2010}]{Wes10}
{Wesson} P.~S.,  2010, \mn@doi [Space Sci. Rev.] {10.1007/s11214-010-9671-x},
  \href {https://ui.adsabs.harvard.edu/abs/2010SSRv..156..239W} {156, 239}

\bibitem[\protect\citeauthoryear{{Wickramasinghe}}{{Wickramasinghe}}{2010}]{Wick2010}
{Wickramasinghe} C.,  2010, \mn@doi [Int. J. Astrobiol.]
  {10.1017/S1473550409990413}, \href
  {https://ui.adsabs.harvard.edu/abs/2010IJAsB...9..119W} {9, 119}

\bibitem[\protect\citeauthoryear{{Wolf} \& {Toon}}{{Wolf} \&
  {Toon}}{2015}]{WT15}
{Wolf} E.~T.,  {Toon} O.~B.,  2015, \mn@doi [J. Geophys. Res. Atmos.]
  {10.1002/2015JD023302}, \href
  {https://ui.adsabs.harvard.edu/abs/2015JGRD..120.5775W} {120, 5775}

\bibitem[\protect\citeauthoryear{{Yule}}{{Yule}}{1924}]{Yu24}
{Yule} G.~U.,  1924, \mn@doi [Phil. Trans. R. Soc. Lond. Ser. B]
  {10.1098/rstb.1925.0002}, 213, 21

\bibitem[\protect\citeauthoryear{{Zhu} et~al.,}{{Zhu} et~al.}{2017}]{ZUN17}
{Zhu} W.,  et~al., 2017, \mn@doi [Astron. J.] {10.3847/1538-3881/aa8ef1}, \href
  {https://ui.adsabs.harvard.edu/abs/2017AJ....154..210Z} {154, 210}

\bibitem[\protect\citeauthoryear{{Zubrin}}{{Zubrin}}{2001}]{Zubrin2001}
{Zubrin} R.,  2001, J. Br. Interplanet. Soc., \href
  {https://ui.adsabs.harvard.edu/abs/2001JBIS...54..262Z} {54, 262}

\makeatother
\end{thebibliography}

\bsp	
\label{lastpage}
\end{document}